\documentclass[letter]{aa} 

\usepackage{epsfig}     
\usepackage{graphicx,color}     
\usepackage{amssymb}            
\usepackage{url}                
\usepackage{amsmath}            
\usepackage{rotating}                   
\usepackage{float}                      
\usepackage{textcomp}           
\usepackage{epstopdf}
\usepackage{dcolumn}
\usepackage{times}
\usepackage{tabularx}
\usepackage{hyperref}
\hypersetup{
    colorlinks,
    citecolor=blue,
    filecolor=blue,
    linkcolor=blue,
    urlcolor=blue,
    menucolor=black}
\usepackage{ulem} 
\usepackage[english]{babel}
\usepackage{booktabs}
\usepackage{gensymb}
\usepackage{subfig}
\usepackage{appendix}

\usepackage{etoolbox}

\makeatletter

\makeatletter
\renewcommand*\aa@pageof{, page \thepage{} of \pageref*{LastPage}}
\makeatother

\begin{document}

\title{Data-driven Magnetohydrodynamic Simulation of the Initiation of a Coronal Mass Ejection with Multiple Stages}

\author{J. H. Guo\inst{1,2}, S. Poedts\inst{1,3}, B. Schmieder\inst{1,4,5}, Y. Guo\inst{2}, C. Zhou\inst{2}, H. Wu\inst{2,1}, Y. W. Ni\inst{2}, Z. Zhong\inst{2}, Y. H. Zhou\inst{2}, S. H. Li\inst{2}, P. F. Chen\inst{2}}
\institute{Centre for Mathematical Plasma Astrophysics, Department of Mathematics, KU Leuven, Celestijnenlaan 200B, B-3001 Leuven, Belgium \\
\email{jinhan.guo@kuleuven.be}
\and School of Astronomy and Space Science, Nanjing University, Nanjing, 210046, People's Republic of China\\
\and Institute of Physics, University of Maria Curie-Skłodowska, Pl.\ Marii Curie-Skłodowskiej 5, 20-031 Lublin, Poland\\
\and LIRA, Observatoire de Paris, CNRS, UPMC, Universit\'{e} Paris Diderot, 5 place Jules Janssen, 92190 Meudon, France \\
\and
LUNEX EMMESI Institut, SBIC, Kapteyn straat 1, Noordwijk2201 BB Netherlands
}

\date{Received 8 December 2025; Accepted 9 February 2026}

\abstract
{Coronal mass ejections (CMEs) are the primary drivers of adverse space-weather events, yet their initiation and onset prediction remain insufficiently understood due to the complexity of the magnetic topology and physical processes in real solar source regions. Here, based on fully observational-data-driven magnetohydrodynamic simulation, we successfully reproduce the initiation of a CME originating from the super active region AR 13663, with only a one-minute time lag between the flare peak in observations and the velocity peak of the rising flux rope in the simulation. Moreover, the eruptive structure exhibits a multi-stage kinematic evolution: an initial slow acceleration, a plateau at a nearly stationary height, and a subsequent impulsive acceleration. These stages  correspond to torus instability, the downward tension force exerted by the overlying toroidal field, and fast magnetic reconnection, respectively. Our results highlight the inherently multistage nature of CME initiation in real events. In configurations with strong overlying toroidal fields, the downward toroidal-field-induced tension force can suppress the rise of the flux rope and produce a plateau phase at a nearly stable height, even when torus instability occurs. In contrast, the subsequent fast magnetic reconnection beneath the flux rope can drive the impulsive eruption more effectively. The close agreement between the observed and simulated peak times over one minute demonstrates the strong potential of our data-driven model for predicting CME onset.}

\authorrunning{Guo et al.}
\titlerunning{Data-driven MHD Simulation of CME Initiation}

\maketitle

\section{Introduction}

Coronal mass ejections (CMEs) are large-scale expulsions of magnetised plasma from the solar corona into interplanetary space, constituting the largest eruptive events currently known in the solar system \citep{Chen2011, Webb2012, Schmieder2015}. Moreover, CMEs are considered the primary drivers of adverse space-weather events, such as geomagnetic storms, which can potentially impact satellite operations and communication and navigation systems \citep{Gosling1993, Schrijver2015}. Consequently, understanding how CMEs are initiated and evolve in the low corona is essential to improving space-weather forecasting capabilities. 

The early-stage kinematics of CMEs are commonly described in terms of a quasi-static phase \citep{Xing2018}, a slow-rise phase \citep{Cheng2020}, and an impulsive phase \citep{Zhang2001, Zhang2006}. The quasi-static and slow-rise phases prior to eruption can be explained by the formation of a current sheet and the response of driving flows \citep{Xingy2024, Liuq2025}, with moderate reconnection occurring at the hyperbolic flux tube \citep[HFT;][]{Titov2002} beneath the flux rope \citep{Cheng2023, Xing2024}. The impulsive phase, characterised by the onset of rapid CME acceleration, can be  triggered mainly  by several mechanisms, e.g., torus instability \citep[TI;][]{Kliem2006, Aulanier2010}, fast tether-cutting magnetic reconnection below the flux rope \citep{Jiang2021}, and breakout reconnection above it \citep{Antiochos1999}. These models establish a basic physical framework for understanding how a CME develops and ultimately erupts.

Although the models described above capture several key physical processes involved in CME initiation (before and at the impulsive phase), their application to forecasting real solar eruptions remains highly challenging. This is because they are based on simplified magnetic configurations (e.g., bipoles or quadrupoles) and on idealised driving flows (e.g., shearing or converging motions), which differ significantly from the conditions in real active regions. The evolution of real active regions involves the interplay of shearing and converging flows, magnetic flux emergence \citep{Chen2000, Fan2018}, and collisional shearing motions between different magnetic systems \citep{Chintzoglou2019}, leading to highly complex and twisted magnetic topologies. As a result, CME kinematics observed in the literature are often multi-staged and highly variable, making real-time CME forecasting particularly difficult.

For this consideration, observational data-driven simulations have recently been developed and have become essential tools for studying CME initiation and forecasting their onset \citep{Jiang2022}. Here, using our recently developed data-driven MHD model \citep{Guo2019, Zhong2023, Guo2024}, we investigate the initiation of a CME in the super active region 13663, which produced more than 20 M-class or higher solar flares within six days. This Letter has two main objectives. First, we aim to assess the capability of our data-driven model to forecast real and complex solar eruptions. Second, we explore the CME onset, which is highly complex in reality and often exhibits multiple stages, and discuss the roles of torus instability and magnetic reconnection. Section~\ref{sec:obs} shows an overview of observations and describes the numerical model. In Section~\ref{sec:res}, we present the simulation results, which are followed by summary and concluding remarks in Section~\ref{sec:sum}.

\section{Overview of Observations and Modeling}\label{sec:obs}

As investigated by \citet{Schmieder2025} and \citet{Lekshmi2025}, NOAA Active Region (AR) 13663 was classified as a Hale-$\beta\gamma\delta$ class and produced more than 20 intense flares and CMEs. Interactions between CMEs originating from ARs 13663 and 13664 subsequently induced a major geomagnetic storm, with a peak $D$st of –412 nT. In this letter, we focus on the X1.3 flare and its associated CME initiation that occurred at approximately 06:00 UT on 2024 May 5.

\begin{figure}[!ht]
    \centering
    \includegraphics[width=8cm]{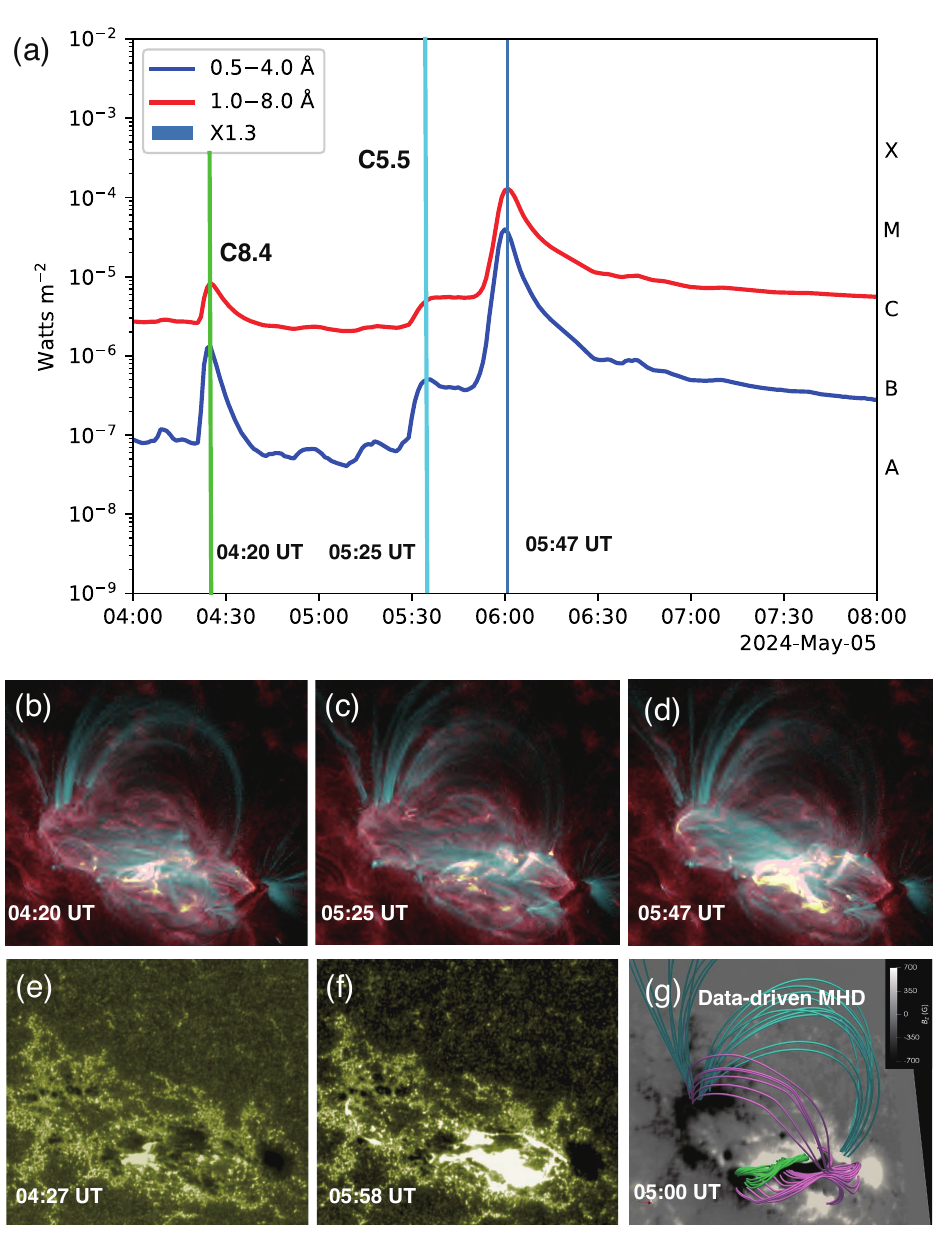}
    \caption{Panel~(a) shows the soft X-ray light curves (red: 1--8~\AA; blue: 0.5--4~\AA) from 04:00 to 08:00~UT on 2024 May~5. The green, sky-blue, and dark-blue vertical lines mark the confined C8.4 and C5.5 flares, and the eruptive X1.3 flare, respectively. Panels~(b)--(d) present composite SDO/AIA 304~\AA\ (red) and 131~\AA\ (blue) images at the peak times of the three flares. Panels~(e) and (f) show the 1600~\AA\ flare-ribbon morphology of the first confined flare and the third eruptive flare, respectively. Panel~(g) displays the 3D magnetic field configuration from the data-driven MHD model at 05:00~UT, where the green, pink, and cyan lines represent the flux rope, the fan--spine structure, and the surrounding coronal loops, respectively.} 
    \label{fig1}
\end{figure}

As shown in Figure~\ref{fig1}a, three flares were detected by the GOES indicated by the soft X-ray light curve between 04:00 and 08:00 UT, including two confined C8.4 and C5.5 flares at $\sim$ 04:20 UT and $\sim$ 05:25 UT, and an eruptive X1.3 flare with an associated faint CME at $\sim$ 05:47 UT. Figures~\ref{fig1}b–d present composite SDO/AIA 304 and 131 \AA\ images at the peak moments of the three flares, while Figures~\ref{fig1}e and f display the corresponding 1600 \AA\ flare ribbons. The eruptive flare exhibits multiple elongated and circular ribbon structures, indicating that the magnetic topology of this active region is highly complex. Figure~\ref{fig1}g shows the 3D magnetic fields at 05:00 UT, from which one can observe a twisted flux rope, fan-spine structure, and overlying coronal loops resembling observations. Appendix~\ref{sec:num} introduces the numerical setup.

\section{Results}\label{sec:res}

\begin{figure}[!ht]
    \centering
    \includegraphics[width=9cm]{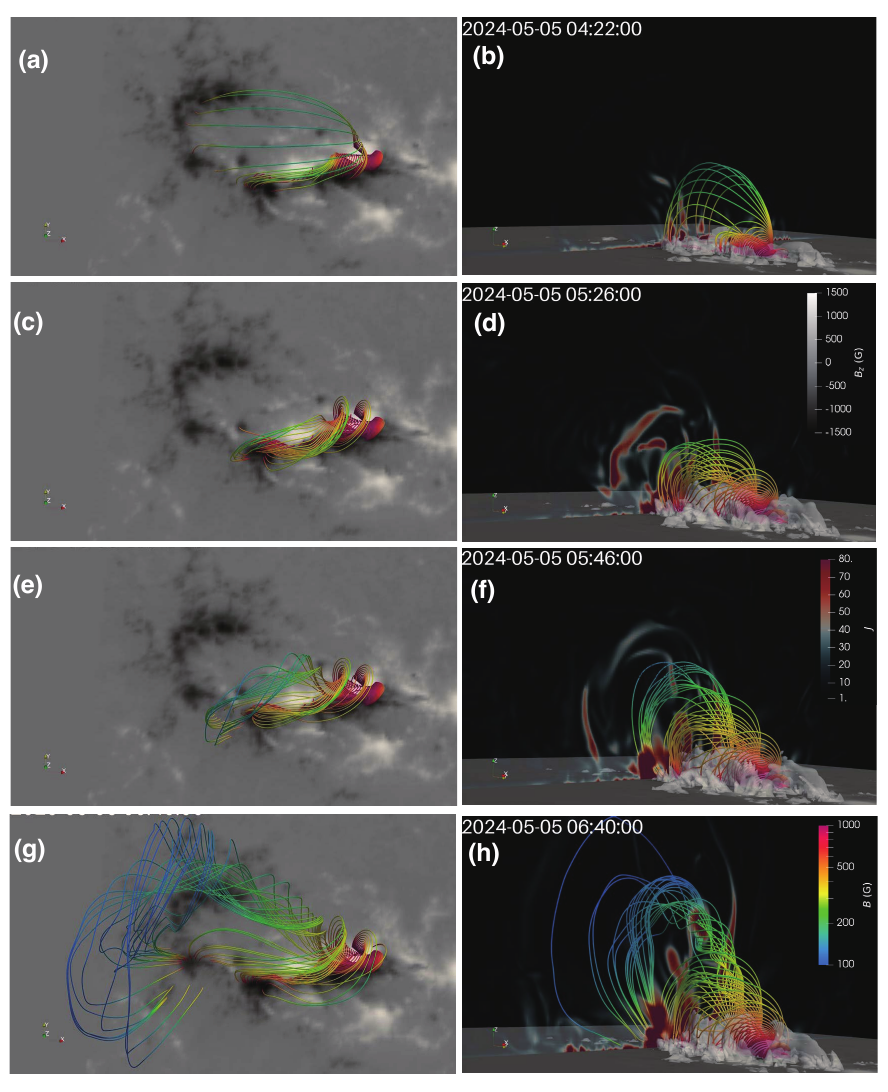}
    \caption{Evolution of the 3D magnetic field lines in the simulation. The left and right panels show the top and side views, respectively. The field lines are colour-coded by magnetic-field strength. The side-plane slice displays the electric current density~$J$, and the grey contours highlight regions of enhanced current. The associated movie is available online.} 
    \label{fig2}
\end{figure}

Figure~\ref{fig2} shows the global evolution of the 3D magnetic field before and during the CME initiation, from which we can see the formation and eruption of a twisted magnetic flux rope. At 04:22 UT (panels a–b), the coronal magnetic field is dominated by highly sheared arcades above the polarity inversion line. As the driving flows persist, a twisted flux rope forms at about 04:46 UT. Thereafter, the flux rope continues to accumulate twist and slowly rises. After 05:40 UT, it undergoes rapid expansion and upward acceleration, and the magnetic system transitions into the eruptive phase at about 05:46 UT. During this stage, the magnetic-field strength of the flux rope decreases significantly, and it develops into a large-scale, strongly twisted structure that ultimately drives the CME. Notably, the CME flux rope in the simulation leans toward the northeast, consistent with the CME trajectory observed near the solar north pole. Moreover, the starting time of the impulsive eruption of the flux rope is nearly the peak time of the X1.3 flare.

\begin{figure}[!ht]
    \centering
    \includegraphics[width=9.5cm]{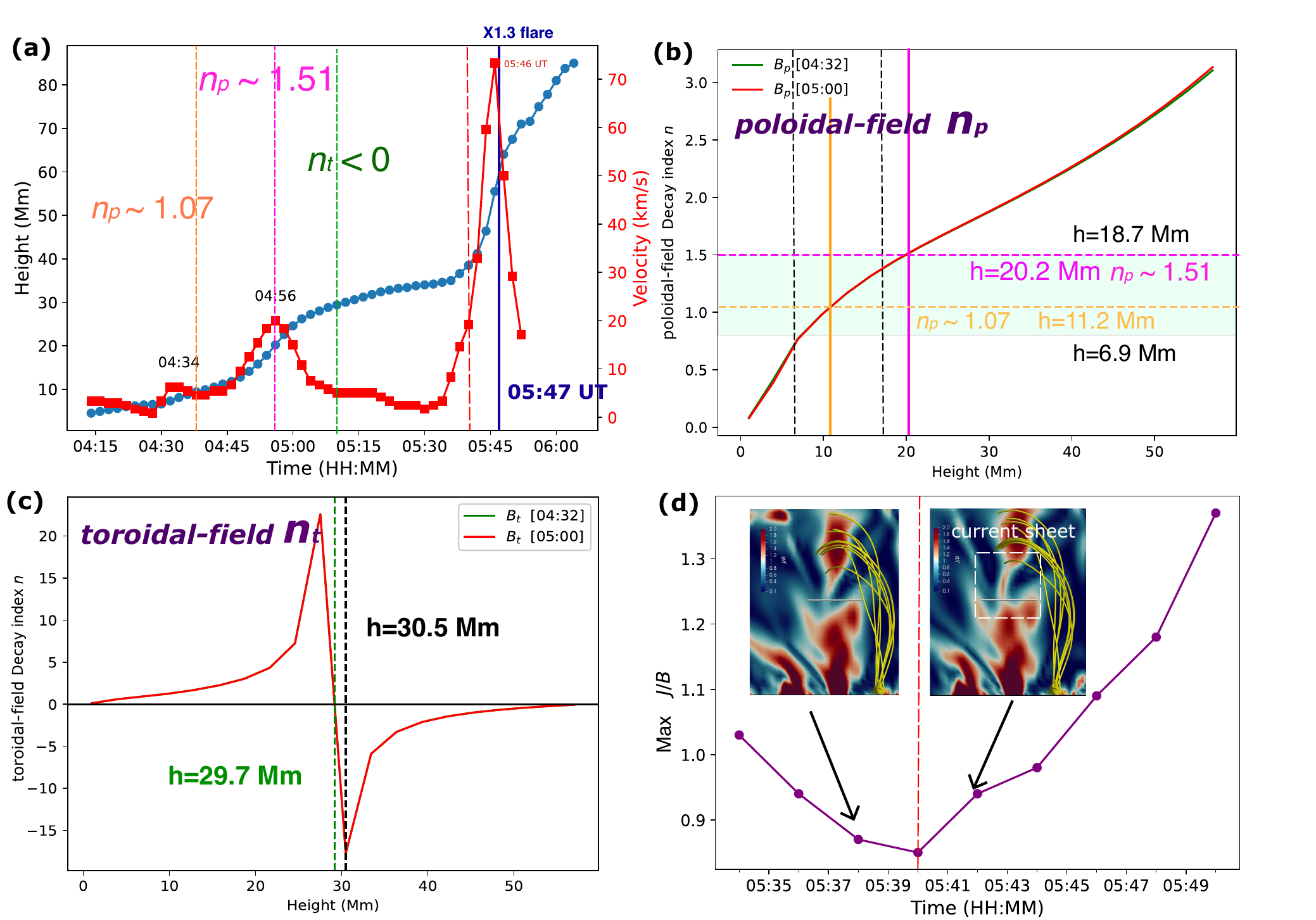}
    \caption{Panel (a) shows the kinematics of the eruptive structures, where the blue and red dots and lines represent the apex height and the derived velocity, respectively. Panel (b) presents the distribution of the decay index computed from the $B_{y}$ component (poloidal field). The cyan-shaded regions indicate the typical threshold for torus instability ($n=0.8$–$1.5$). The green and red lines show the results derived from the potential field at 04:32 and 05:00~UT, respectively. The orange and pink lines mark the critical height and corresponding decay index at the onset (04:46~UT) and peak (04:56~UT) of the second acceleration phase. Panel (c) shows the decay-index distribution computed from the $B_{x}$ component (toroidal field), with the green and black dashed vertical lines indicating the heights where the decay index equals zero and reaches its minimum, respectively. Panel (d) shows the evolution of the maximum $J/B$ along the horizontal line in the inset $J/B$ images. The red dashed vertical line indicates the time when the maximum $J/B$ begins to increase.} 
    \label{fig3}
\end{figure}

To track the kinematics of the erupting flux rope, we calculate the apex height (blue dots and curve) of the eruptive structure and derive its rising speed (red dots and curve), as shown in Figure~\ref{fig3}a. Similar to the soft X-ray light curve, the flux-rope kinematics exhibit three distinct acceleration episodes. The accelerations begin at 04:30 UT, 04:46 UT, and 05:34 UT and reach their respective peaks at 04:34 UT, 04:56 UT, and 05:46 UT. This indicates that the CME onset proceeds through multiple kinetic stages. Moreover, two notable features are found in the velocity evolution of the flux rope: (1) the third velocity peak is approximately 3.5 times larger than the second peak; and (2) a decrease in speed occurs between the second and third peaks, in which the flux rope almost stops at a stable height. It is also worth noting that the time lag between the X1.3 flare peak and the third velocity increase is only about 1 minute, suggesting that our data-driven simulation successfully reproduces the timing and characteristics of the observed solar eruption.

At the time of the first acceleration, the twisted flux rope had not yet fully formed. Nevertheless, magnetic reconnection within the fan–spine structure above the sheared arcade could already have occurred, initiating the initial rise of the eruptive structures and facilitating the subsequent formation of the flux rope around 04:46 UT. Additionally, the quasi-separatrix layers \citep[QSLs;][]{Priest1995, demo1996} associated with the fan–spine topology agree closely with the morphology of the C8.4 flare that peaked at 04:20 UT (Figure~\ref{fig1}e). Additional details on the evolution of the QSLs and the photospheric twist distribution are provided in Figure~\ref{ap1} of the Appendix~\ref{sec:QSL}. These findings suggest that the first acceleration phase is associated with the C8.4 confined flare driven by fan–spine reconnection.

Afterwards, a twisted flux rope forms at approximately 04:46 UT and begins to rise slowly. At this moment, the apex height of the flux rope reaches approximately 11.2~Mm. To investigate the role of torus instability in its subsequent acceleration, we calculate the decay index at 04:32~UT (green curves) and 05:00~UT (red curves), respectively, to minimise the influence of the bottom-driven boundaries on the decay-index profile. Figure~\ref{fig3}b shows the poloidal-field decay index $n_{p}$ computed from the poloidal magnetic field component ($B_p$). Further details on the computation methods and physical meanings of the decay indices are provided in Appendix~\ref{decay index}. 

It should be noted that the critical threshold of torus instability is not a fixed value but can vary between $n\approx0.8$ and $1.5$ according to previous theoretical analyses \citep{Demoulin2010} and MHD simulations \citep{Zuccarello2015}. Thus, torus instability is expected to occur within the range, with corresponding critical heights of 6.9~Mm for $n=0.8$ and 18.7~Mm for $n=1.5$ (cyan band in Figure~\ref{fig3}b). At 04:46~UT, coinciding with the onset of the second acceleration phase, the flux-rope apex reaches 11.2~Mm, where the decay index is about 1.07. By 04:56~UT, when the rising velocity peaks, the apex height increases to approximately 20.2~Mm, corresponding to a decay index of 1.51, which is close to the commonly adopted critical value of 1.5 \citep{Aulanier2010}. These results suggest that torus instability likely contributes to the second acceleration stage of the flux rope, with an effective instability threshold in the range $n\approx$1.07--1.51, during which the torus instability contributes to the rising velocity.

After the second acceleration phase, the flux rope rises to a plateau from 05:04~UT to 05:36~UT, even though the torus instability is likely already at work. Such a plateau is absent in previous MHD simulations based on simple bipolar configurations \citep{Aulanier2010}. A key difference between our model and the idealised bipole is the presence of strong toroidal magnetic fields (Figure~\ref{ap3}a), which may induce downward force \citep{Myers2015}. In Figure~\ref{fig3}c, we compute the toroidal-field decay index $n_{t}$ based on $B_{x}$, which is approximately aligned with the flux-rope axis. The results show that toroidal-field decay index $n_{t}$ becomes negative at a height of 29.7~Mm and reaches its minimum around 30.5~Mm. This indicates that the toroidal-field-induced tension force becomes increasingly dominant at these heights, naturally explaining why the flux rope remains nearly stationary at  heights between 30 and 35~Mm during 05:04–05:36~UT. The gradual displacement of the flux rope during this stage could be attributed to moderate magnetic reconnection \citep{Inoue2018}, photospheric driving flows \citep{Liuq2025} or plasma inertia.

After 05:36~UT, the rising velocity of the flux rope increases again and undergoes significant acceleration around 05:40~UT, when the flux rope apex reaches a height of approximately 38.7~Mm. At this height, the toroidal-field decay index $|n_{t}|$ (still negative) has diminished to nearly one-tenth of its minimum, suggesting that the overlying toroidal field no longer strengthens rapidly with height. This implies that the downward tension force exerted by the toroidal fields persists to some extent but does not continue to significantly increase with height. The rising speed peaks at 05:46~UT, reaching a value about 3.5 times larger than that of the second velocity peak. To investigate the reasons for this behaviour, we examine the temporal evolution of the maximum $J/B$ below the erupting flux rope (Figure~\ref{fig3}d). First, the maximum $J/B$ decreases due to the passage of the rising flux rope. Then, it increases sharply beginning at 05:40~UT, suggesting the onset of fast magnetic reconnection beneath the flux rope, which drives its impulsive upward motion \citep{Jiang2021}.

\section{Summary and Conclusion}\label{sec:sum}

Based on data-driven MHD simulations, we investigate the initiation of the CME associated with an X1.3 solar flare in NOAA active region 13663. The time lag between the observed flare peak and the velocity peak in the simulation is only one minute, demonstrating that our data-driven model closely reproduces observations. Moreover, the simulation reveals multiple stages in the kinematics of the eruptive structure during CME initiation. These results indicate that our data-driven approach is fundamentally capable of modelling CME onset, and can serve as an effective tool to understand the physics behind complex observations.

The simulation results indicate that in a magnetic configuration with strong overlying toroidal fields, the kinematics of CME precursors likely evolve through multiple stages: an initial slow acceleration, a plateau at an approximately stable height, and a subsequent impulsive acceleration (Figure~\ref{fig3}a and \ref{ap3}). These stages are associated with torus instability \citep{Kliem2006, Aulanier2010}, the downward tension force of the overlying toroidal field \citep{Myers2015, Zhong2021, Guo2024, Zhang2024}, and rapid magnetic reconnection \citep{Jiang2021}, respectively. The presence of the plateau suggests that the strong overlying toroidal field can temporarily suppress the eruption. More importantly, the simulation demonstrates that magnetic reconnection acts as the primary trigger of the solar eruption. The appearance of the plateau further implies that, although torus instability may occur, it does not necessarily lead to a drastic eruption. This indicates that the external magnetic configuration strongly influences CME initiation. As shown in \citet{Kliem2011}, a saddle-like profile of the poloidal-field decay index $n_{p}$ can lead to a confined flare followed by a successful eruption. In this case, however, $n_{p}$ increases monotonically, while the toroidal-field decay index $n_{t}$ becomess negative above a certain height, where enhanced toroidal tension leads to the plateau stage. The multiple stages during the CME initiation are also found in \citet{Zhou2006}, in which the kink instability and filament drainage contribute to the first acceleration, and the reconnection leads to the eruption. These studies indicate that the impulsive phase is governed by fast magnetic reconnection, whereas the preceding moderate acceleration can arise from a variety of mechanisms.

Additionally, the time difference between the observed flare peak and the velocity peak of the flux rope in our simulations is only about 1 minute, highlighting the close agreement between the model and observations. Overall, the results presented in this paper demonstrate the strong capability of numerical data-driven modelling in accurately reproducing the onset and early dynamics of CMEs, providing a powerful tool for understanding and predicting solar eruptions.

\begin{acknowledgements}
J.H.G., P.F.C., and Y.G. are supported by the National Key R\&D Program of China 2020YFC2201200, 2022YFF0503004, NSFC (12503063, 12333009, 12127901) and the China National Postdoctoral Program for Innovative Talents fellowship under Grant Number BX20240159. SP is funded by the European Union. However, the views and opinions expressed are those of the author(s) only and do not necessarily reflect those of the European Union or ERCEA. Neither the European Union nor the granting authority can be held responsible. His project (Open SESAME) has received funding under the Horizon Europe program (ERC-AdG agreement No 101141362). These results were also obtained in the framework of the projects C16/24/010 C1 project Internal Funds of KU Leuven), G0B5823N and G002523N (WEAVE) (FWO-Vlaanderen), and 4000145223 (SIDC Data Exploitation (SIDEX2), ESA Prodex). The numerical calculations in this paper were performed in the cluster system of the High Performance Computing Center (HPCC) of Nanjing University. 
\end{acknowledgements}

\bibliography{bibtex}{}
\bibliographystyle{aa}

\begin{appendix}

\section{Numerical Setup of Data-driven MHD Modeling}  \label{sec:num}
To study the initiation process of the CME, we adopt our data-driven MHD model to simulate the evolution of this active region from 04:12 UT to 07:00 UT. The governing equations are as follows:

\begin{eqnarray}
 && \frac{\partial \rho}{\partial t} +\nabla \cdot(\rho \boldsymbol{v})=0,\label{eq1}\\
 && \frac{\partial (\rho \boldsymbol{v})}{\partial t}+\nabla \cdot(\rho \boldsymbol{vv}-\boldsymbol{BB})+\nabla (\frac{\boldsymbol{B}^2}{2})=0,\label{eq2}\\
 && \frac{\partial \boldsymbol{B}}{\partial t} + \nabla \cdot(\boldsymbol{vB-Bv})=0,\label{eq3}
\end{eqnarray}
where $\boldsymbol{B}$, $\boldsymbol{v}$ and $\rho$ denote magnetic field, velocity and density, respectively. The initial condition for magnetic fields is provided by the potential-field model with the Green's function \citep{Chiu1977} at 04:12 UT. The density is provided by the hydrostatic solution of a stratified atmosphere from the chromosphere to the corona \citep{Guo2023a, Guo2023b, Guo2024}. Regarding the driven boundary, we adopt the SHARP vector magnetic fields and the derived velocities with the Differential Affine Velocity Estimator for Vector Magnetograms \citep[DAVE4VM;][]{Schuck2008} method, implementing the $\mathbf{v}$\&$\mathbf{B}$-driven boundary \citep{Guoy2024}. They are provided in the cell centre of the inner ghost cell near the physical computational domain, and those in the outer ghost cells are obtained via equal-gradient extrapolation. In particular, to keep the simulation environment closer to observations, we directly use the observational magnetograms and derived velocities without pre-processing, with maximum magnetic fields exceeding 2000 G and driven velocities below 1~km s$^{-1}$. This approach ensures that the simulation results are directly comparable to real observations, capturing the physical evolution of the eruptive structure with high fidelity. As a result, the evolution of coronal magnetic fields is fully governed by photospheric flows in observations, and the formation and eruption of the flux rope are self-consistent.

The partial differential equations are numerically solved using the Message Passing Interface Adaptive Mesh Refinement Versatile Advection Code \citep[MPI-AMRVAC\footnote{http://amrvac.org},][]{Xia2018, Keppens2023}. The computational domain is $[x_{min},x_{max}]\ \times [y_{min},y_{max}]\ \times [z_{min},z_{max}] = [-172.3, 172.3]\ \times [-172.3,172.3]\ \times[1,345.7]\;\rm Mm^{3}$, with an effective mesh grid of $480 \times 480 \times 480$, adopting a four-level adaptive mesh refinement. In particular, as in \citet{Jiang2021} and \citet{Wu2025}, to refine the current sheet in the simulation, an additional AMR criterion is imposed, $J\triangle /B>0.2$, where $J$ is the current density and $\triangle$ is the spatial grid size.

\section{Distributions of QSLs and Twist, and Comparison with Observations}  \label{sec:QSL}
\begin{figure*}[!ht]
    \sidecaption
    \includegraphics[width=12.5cm]{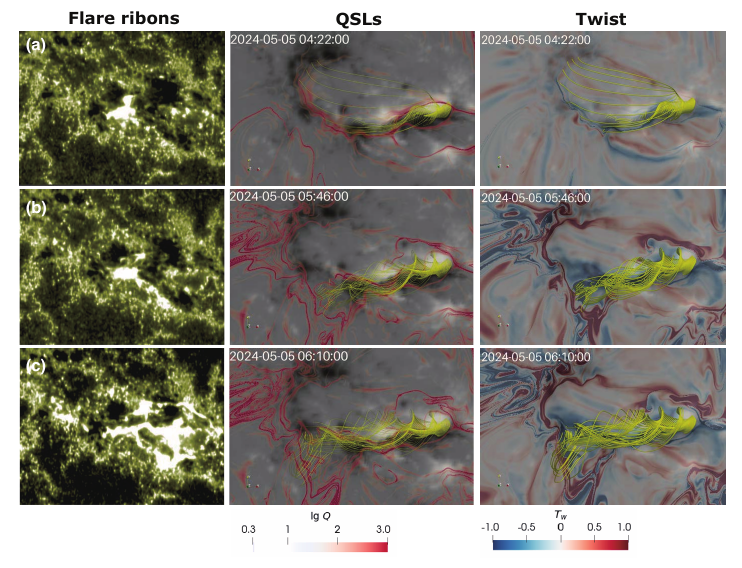}
    \caption{Comparison of flare ribbons in 1600 \AA, and the distributions of QSLs and twist on the bottom. The red contours in the middle panels denote the regions with lg $Q>2$. The blue and red regions in the right panel denote negative and positive twists, respectively. The yellow lines represent the traced eruptive magnetic structures. }
    \label{ap1}
\end{figure*}

To compare the simulation results with observations, we compute the distributions of QSLs and the twist number on the bottom plane using the open-source code implemented by \citet{Liu2016}, as shown in Figure~\ref{ap1}. It is found that QSLs closely resemble the morphology of flare ribbons, indicating that the magnetic topology in the simulation results is comparable to that observed. Additionally, it is observed that the accumulation of twist during the CME buildup and the eastern slip of the east footpoint are denoted by high-twist-number regions.

Moreover, although the onset times of the three flare peaks in the observations and the corresponding velocity peaks in the simulations are not perfectly aligned, we still infer that they are physically related. For the first confined flare, the flux rope has not yet formed; the flare ribbons arise from fan–spine magnetic reconnection. The second confined flare is associated with the flux rope rising due to torus instability, thereby enhancing reconnection above the null point. The third flare is more complex, showing multiple ribbons produced by rapid magnetic reconnection that ultimately trigger the eruption.

\section{Typical Magnetic Topology}

\begin{figure}[!ht]
    \centering
    \includegraphics[width=9cm]{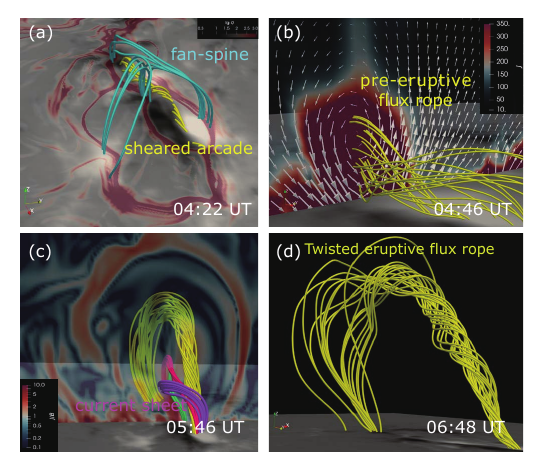}
    \caption{Typical magnetic topologies involved in CME initiation. Panels (a), (b), (c), and (d) show, respectively, the fan–spine structure and sheared arcades at 04:22 UT, the pre-eruptive flux rope at 04:46 UT, the current sheet at 05:46 UT, and the twisted eruptive flux rope at 06:48 UT.} 
    \label{ap2}
\end{figure}

Here, we present the characteristic magnetic topology during the CME initiation. Figure \ref{ap2}a shows the sheared arcades and the overlying fan–spine structure at 04:22 UT, where the fan–spine reconnection triggers the C8.4 confined flare. Figure \ref{ap2}b displays the pre-eruptive twisted flux rope at 04:46 UT. The flux rope contains an inverse magnetic-field component and remains anchored to the photosphere, where the second acceleration phase begins. Figure \ref{ap2}c illustrates the erupting flux rope together with the current sheet beneath it, in which the purple and green lines represent two groups of sheared arcades prior to the eruption, while the pink lines denote the newly formed field lines after magnetic reconnection. The magnetic reconnection drives the eruption into the impulsive phase. First, as discussed in \citet{Jiang2021} and \citet{Jiang2024}, the slingshot effect of the newly reconnected field lines (pink field lines) can efficiently accelerate the eruption. Second, the downward Lorentz force is weak at this height: the poloidal-field decay index $n_{p}$ exceeds 1.5, indicating that the strapping field decreases sufficiently rapidly with height to constrain the eruption; and the toroidal-field decay index $n_{t}$ increases from $\sim -20$ to nearly 0, implying that the toroidal-field-induced downward tension force no longer strengthens significantly with height. Consequently, once magnetic reconnection sets in, the eruption can be triggered and accelerated more effectively under reduced constrains. In particular, the purple field lines connect to the eastward negative-polarity sunspot, suggesting that the eastward slipping motion of the eastern flux-rope footpoint could be related to the impulsive magnetic reconnection. Figure \ref{ap2}d shows the twisted eruptive flux rope at 06:48 UT, revealing that its western leg above the compact PIL is more highly twisted than the eastern leg rooted in diffused magnetic polarities. In addition, rotational velocity patterns are absent in the velocity-field distribution at the flux-rope footpoints, indicating that the twist does not result from rigid twisting motions in the photosphere. This implies that the enhanced twist is primarily generated through magnetic reconnection during the eruption.

\section{Decay Index Derived from Different Magnetic Field Components}\label{decay index}

To quantify the influence of external magnetic fields on the triggering and evolution of the solar eruption, we calculate the decay index using the poloidal ($B_{p}$), toroidal ($B_{t}$), and total horizontal magnetic fields ($B_{h}$). The decay index is defined as
\begin{equation}
n_{i}(z) = -\frac{d \log B_{i}}{d \log z} = -\frac{z}{B_{i}}\frac{dB_{i}}{dz},
\end{equation}
where $i = p,\, t,\, h$ corresponds to the poloidal, toroidal, and total horizontal components, respectively. The resulting quantities are referred to as the poloidal-field decay index ($n_{p}$), toroidal-field decay index ($n_{t}$), and total horizontal-field decay index ($n_{h}$). Because the pre-eruption flux rope is oriented approximately along the $x$-direction, $B_{y}$ and $B_{x}$ are used as reasonable proxies for the poloidal and toroidal components in the flux-rope coordinate system.

Figure~\ref{ap3} shows the profiles of $n_{p}$, $n_{t}$, and $n_{h}$. These profiles exhibit distinctly different shapes: $n_{p}$ shows a monotonically increasing trend, $n_{t}$ displays a dipolar shape (two opposite-sign peaks), and $n_{h}$ presents a unimodal (single-peak) shape. Two features are noteworthy. First, the critical height at which the decay index reaches the threshold value of 1.5 differs among the three profiles, while the second acceleration phase in the simulation occurs when $n_{p} \sim 1.5$. This highlights the importance of using the poloidal field to compute the decay index relevant to torus instability, rather than $n_{t}$ and $n_{h}$. Second, the height of the dip in $n_{t}$ nearly coincides with that of the maximum in $n_{h}$, which likely corresponds to the height of the plateau stage during the eruption. As a result, based on this data-driven MHD simulation, we suggest that the poloidal-field decay index ($n_{p}$) governs the onset of the solar eruption through the torus instability (the slow-acceleration phase), whereas the toroidal-field decay index ($n_{t}$) controls the subsequent evolution (the plateau stage).

This indicates that the initial external magnetic configuration has a significant impact on CME initiation. For instance, as shown by \citet{Guo2010} and \citet{Kliem2011}, a saddle-like profile (a rise–fall–rise pattern) of the poloidal-field decay index $n_{p}$ can lead first to a confined flare and then to a subsequent eruptive CME. The initial confinement occurs because the poloidal-field–induced strapping force dominates within the torus-unstable regime at the $n_{p}$ minimum. In contrast, in the present event, $n_{p}$ exhibits a monotonically increasing trend, whereas the toroidal-field decay index $n_{t}$ displays a dipolar profile. The region of negative $n_{t}$ corresponds to a locally strengthening toroidal field, which enhances the toroidal-field–induced tension force, thereby leading to the  plateau phase in the CME initiation.

\begin{figure}[!ht]
    \centering
    \includegraphics[width=9.5cm]{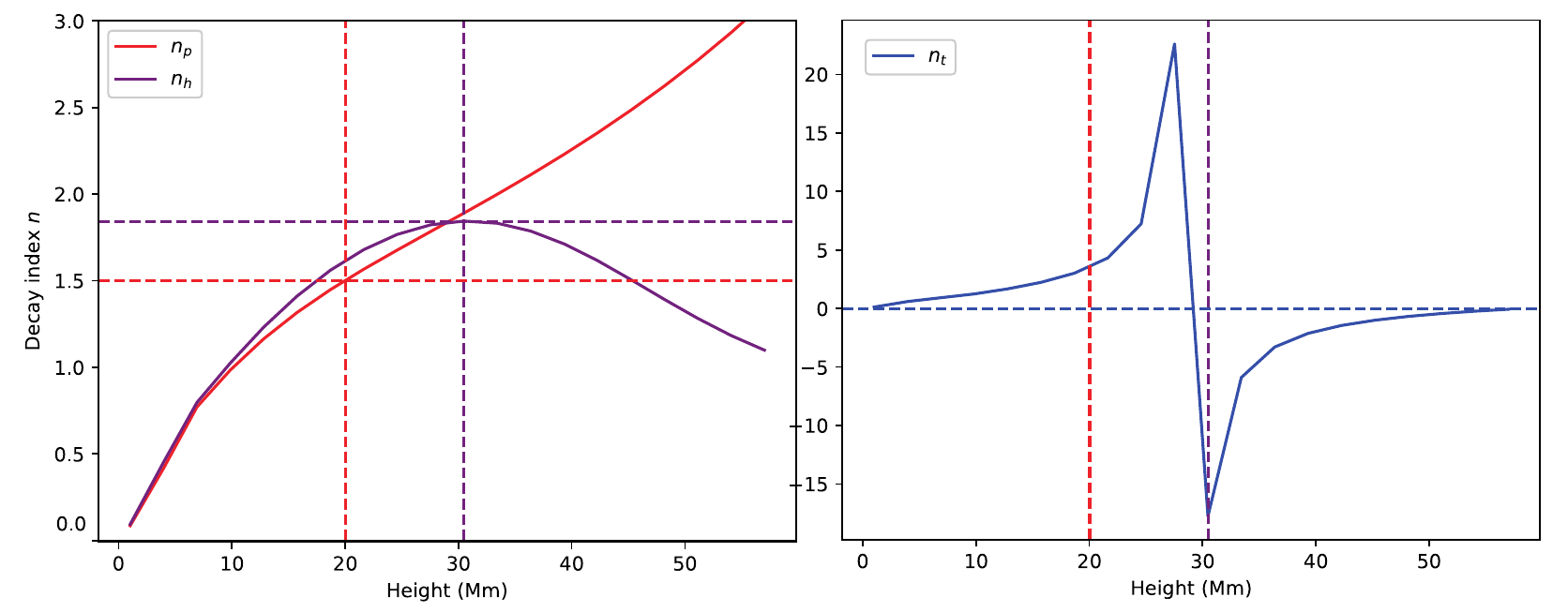}
    \caption{Profiles of the poloidal-field decay index ($n_{p}$; red), toroidal-field decay index ($n_{t}$; blue), and total horizontal-field decay index ($n_{h}$; purple), derived from the corresponding poloidal ($B_{p}$), toroidal ($B_{t}$), and total horizontal ($B_{h}$) magnetic fields. The red and purple vertical lines indicate the heights of $n_{p}=1.5$ and the maximum $n_{h}=1.84$, respectively.} 
    \label{ap3}
\end{figure}

\section{Multi-stage Kinematics of CME Precursors in Magnetic Configurations with Strong Overlying Toroidal Fields}

Figure~\ref{ap3} shows the magnetic configuration and the corresponding time–distance diagram of the CME precursors (for example, filament and hot channels) extracted from our data-driven simulation after flux-rope formation. The simulation results indicate that in the magnetic configuration with strong overlying toroidal magnetic fields, the kinematics of CME precursors likely proceed through multiple stages: an initial slow acceleration, a plateau at a nearly stable height, and a subsequent impulsive acceleration (Figure~\ref{ap3}b). 

Based on the analysis in this study and previous works, we propose that the initial slow acceleration corresponds to the slow-rise phase observed in bipole configurations, which can be explained by the direct responses to the current-sheet formation \citep{Xingy2024, Liuq2025}, HFT reconnection \citep{Cheng2023, Xing2024}, and torus instability \citep{Xing2024}. In our simulations, the complexity of the photospheric flows makes it difficult to isolate the contribution of each effect individually. We speculate that these effects likely work in combination. Hereafter, in contrast to bipole configurations, where torus-instability-induced acceleration can directly trigger eruptions \citep{Aulanier2010}, or facilitate current-sheet formation beneath the flux rope and thereby initiate the impulsive phase \citep{Xing2024}. In our case, the presence of strong overlying toroidal fields generates a downward tension force that suppresses further rising \citep{Myers2015, Guo2024}, resulting in a plateau stage. The subsequent impulsive acceleration arises from fast magnetic reconnection beneath the flux rope \citep{Jiang2021}, indicating that fast reconnection is the dominant trigger and driver of the eruption. Consistent with the conclusion of \citet{Liuq2024}, the torus instability may not necessarily trigger an eruption even in the presence of a pre-existing flux rope. This simulation demonstrates that strong overlying toroidal fields can extend the duration of the buildup phase before the impulsive eruption, thereby improving the prediction of the onset time in real events.

\begin{figure}[!ht]
    \includegraphics[width=8cm]{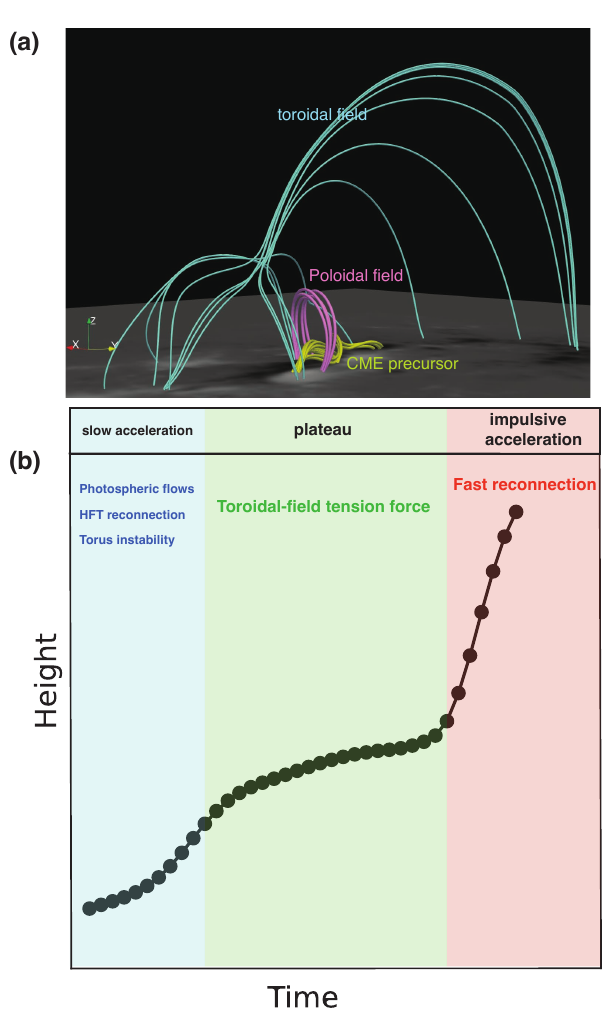}
    \centering
    \caption{Magnetic configuration (a) and the corresponding kinematics of the CME precursors (b) during the initiation process. In panel (a), the yellow, pink, and cyan lines represent the CME precursor (flux rope), the overlying poloidal fields, and the toroidal fields associated with the fan–spine structure, respectively. In panel (b), the blue, yellow, and red bands indicate the three stages of the CME precursors. The black line with black dots shows the time–distance diagram of the CME precursors.} 
    \label{ap3}
\end{figure}

\end{appendix}

\end{document}